\documentstyle[12pt]{article}
\begin{document}
\vspace*{-2cm}

\normalsize
\hspace*{10cm}\parbox{15cm}{IU/NTC 96-01}                       \\
\hspace*{10cm}\parbox{15cm}{ADP-96-2/T207}                     \\

\Large
\begin{center}
Kaon-Nucleus Drell-Yan Processes and Kaon Structure Functions \\
\end{center}

\vspace*{0.4cm}

\normalsize
\begin{center}
{\large J.T. Londergan}                                    \\
{\small\it Department of Physics and Nuclear Theory Center      \\
Indiana University                                              \\
Bloomington IN 47408 USA},                                      \\

\vspace*{0.4cm}
{\large G.Q. Liu and A.W. Thomas}           \\
{\small\it Department of Physics and Mathematical Physics       \\
University of Adelaide                                          \\
Adelaide, S.A., 5005, Australia}

\end{center}

\large
\begin{center}
Abstract
\end{center}
\normalsize

\hspace*{-0.5cm}
We investigate the information which could be obtained from Drell-Yan 
processes with sufficiently intense beams of charged kaons on isoscalar
targets. It is found that 
combinations of $K^+$ and $K^-$ Drell-Yan measurements on isoscalar
nuclear targets would allow one to extract the kaon sea quark distributions. 
These cross sections are also sensitive to the strange valence
quark distribution in the kaon, although one would need a significant
increase over the presently available kaon
fluxes in order to extract this quantity with sufficient accuracy. 

\vspace*{0.5cm}

PACS: 13.60.Hb; 12.40.Gg; 12.40.Vv.

\vspace*{1cm}
{\LARGE -----------------}\\
\hspace*{0.5cm}
{\small e-mail:\\ 
\hspace*{2.0cm}\parbox{15cm}{londergan@iucf.indiana.edu\\
gliu,$\:$ athomas@physics.adelaide.edu.au
}}

\newpage

In principle, the
Drell-Yan [DY] process allows one to separate the valence and sea quark
distributions for nucleons. One can also obtain meson structure functions
by using a combination of Drell-Yan processes and prompt photon data 
[1-6].  
Existing measurements give reasonable constraints on 
pion valence quark distributions, although the lack of sufficient data
at small $x_\pi$ does not allow the pion sea quark distribution to be
extracted.  In a previous letter \cite{llrt} we showed that a comparison
of Drell-Yan processes induced by oppositely charged pions on an 
isoscalar target would allow one to determine the pion
sea quark distribution unambiguously.  

Given a sufficiently intense beam of kaons, one could repeat these
measurements to extract information on the parton distributions in the kaon.
Amongst the many questions of theoretical interest which could be
addressed we mention the difference in the valence distributions of the
strange and non-strange quarks in the kaon, and the possibility of comparing
valence distributions in the kaon and pion. In addition, at the present
time there is no information at all on the quark sea in the kaon.
Our aim is to derive the relevant Drell-Yan cross sections and investigate 
their sensitivity to these aspects of the structure of the kaon.

In the Drell-Yan process\ \cite{drell}, a quark (antiquark) of a certain flavor 
in the projectile annihilates an antiquark (quark) of the same flavor in the 
target, producing a virtual photon which later decays into a 
$\mu^+ -\mu^-$ pair.  For nucleon-nucleus events, the antiquark is necessarily 
part of the sea, so all Drell-Yan events involve at least one sea quark. Since 
a kaon has one valence quark and one valence antiquark, various combinations 
of valence and sea quarks are possible for kaon-nucleus Drell-Yan processes.  
Combinations of positively and negatively charged kaons on isoscalar 
targets allow the valence-valence interactions to be separated in a 
straightforward way.  

Assuming, for illustrative purposes, a deuteron [D] target, the 
Drell-Yan cross sections for positive and negative kaons have the form 
%
\begin{eqnarray}
\sigma_{DY}^{K^+ D}(x,x_K) &=&   
   {2\over 9}\left[ 4 u^{K^+}_{\rm v}(x_K) + 
  \kappa\bar{s}^{K^+}_{\rm v}(x_K) + 
   (10 + 2\kappa)q^K_{\rm s}(x_K) \right]q^N_{\rm s}(x)  \nonumber \\
   &+& {10\over 9}q^K_{\rm s}(x_K)q^N_{\rm v}(x),  \nonumber \\
\sigma_{DY}^{K^- D}(x,x_K) &=& {2\over 9}\left[
   4\bar{u}^{K^-}_{\rm v}(x_K) + \kappa s^K_{\rm v}(x_K) + 
   (10 + 2\kappa)q^K_{\rm s}(x_K)\right] q^N_{\rm s}(x) \nonumber \\ 
   &+& \left[ {8\over 9}\bar{u}^{K^-}_{\rm v}(x_K)
   + {10\over 9}q^K_{\rm s}(x_K)\right] q^N_{\rm v}(x).
\label{eq:Rpm}
\end{eqnarray}
\normalsize
In Eq.\ \ref{eq:Rpm} we assumed the validity of charge symmetry, i.e.\ 
$d^n(x)=u^p(x)$,
$u^n(x)=d^p(x)$, etc. While there have been predictions of significant
charge symmetry violation in the parton distributions of the 
nucleon \cite{rtl,sather}, and the Drell-Yan process may be extremely 
useful in testing those predictions \cite{llrt,ptwo}, any reasonable 
deviation from charge symmetry would have a
negligible effect on the reactions considered here. Assuming charge
symmetry we can write all processes in
terms of parton distributions in the proton, where we defined the 
valence distribution $q^N_{\rm v}(x) \equiv (u^p_{\rm v}(x) + 
d^p_{\rm v}(x))/2$. In addition, for the kaon sea we assumed SU(3) 
symmetry -- i.e. we assumed the up, down and strange quark sea distributions 
in the kaon to be equal ($u_{\rm s}^K=d_{\rm s}^K=s^K_{\rm s}\equiv 
q_{\rm s}^K$) -- since no other information is available on the kaon sea. 
Finally, for the nucleon we assumed 
$q^N_{\rm s}(x) \equiv (u_{\rm s}(x) + d_{\rm s}(x))/2 = s(x)/\kappa$, 
where the fraction $\kappa$ of the strange sea is chosen
to reproduce the experimental ratio of dimuon events to single-muon
events in neutrino-induced reactions \cite{ccfrt,ccfro,cdhsw}   
at the scale $Q^2 = Q_0^2 = 4$ GeV$^2$.  
Under these assumptions, $K^+$ and $K^-$ DY processes on isoscalar 
targets should produce identical cross sections, except for the additional
contribution in the $K^-$ case which is proportional to the 
product of the $\bar{u}$ valence quark in the kaon times the up quark 
distributions in the proton and 
neutron.  Consequently, if we form the linear combination
%
\begin{eqnarray}
\Sigma_{\rm v}^{K D}(x,x_K)  &=& \sigma_{DY}^{K^- D} - \sigma_{DY}^{K^+ D} = 
    {8\over 9}\,\bar{u}^{K^-}_{\rm v}(x_K)\, q^N_{\rm v}(x),
\label{eq:sigmav}
\end{eqnarray}
\normalsize
the $K^+ D$ DY cross section contains no valence-valence contribution 
(it contains only sea-valence and sea-sea terms), while $\Sigma^{KD}_{\rm v}$ 
contains only a valence-valence term (note that our equations 
are correct for any isoscalar nuclear target; however, the quantity 
$q^N_{\rm v}(x)$ would then be the nuclear valence quark distribution 
function.  To extract the free nucleon quark distribution one would
have to understand binding and EMC-type nuclear corrections to the
quark distributions).  

This means that if one measures $K^+$ and $K^-$ Drell-Yan cross sections
on an isoscalar nucleus, then the $K^+D$ cross sections are sensitive to 
sea-valence
and sea-sea interference terms, whereas the valence-valence contribution
can be obtained from $\Sigma^{KD}_{\rm v} = \sigma_{DY}^{K^- D} - 
\sigma_{DY}^{K^+ D}$.  Furthermore, since the quantity $\Sigma^{KD}_{\rm v}$
separates in the variables $x$ and $x_K$, the up quark valence distribution
in the kaon can be obtained by integrating this quantity over all $x$, 
i.e. 
\begin{eqnarray}
\bar{u}_{\rm v}^{K^-}(x_K) &=& {3\over 4} \int_0^1 \Sigma^{KD}_{\rm v}
 (x,x_K)\, dx \quad .
\end{eqnarray}
As a consequence, it may be possible to obtain the nonstrange valence 
quark distribution in the kaon even with reasonably low kaon fluxes.  

If one defines the sea-to-valence ratios for kaon and nucleon, 
%
\begin{eqnarray}
r_{\rm s/v}^N(x)    &\equiv& [ u^p_{\rm s}(x) + d^p_{\rm s}(x) ] /[
  u^p_{\rm v}(x) + d^p_{\rm v}(x) 
  ],  \nonumber \\
q_{\rm s/v}^K(x)  &\equiv&  q^K_{\rm s}(x)/u^K_{\rm v}(x) , \nonumber \\  
R_{\rm s/u}^K(x)  &\equiv&  s^K_{\rm v}(x)/u^K_{\rm v}(x),.  
\label{eq:rsvdef}
\end{eqnarray}
\normalsize
we find the following ratio 
%
\begin{eqnarray}
R^K_{\rm s/v}(x, x_K)  &\equiv& \sigma_{DY}^{K^+ D} / \Sigma_{\rm v}^{K D}  
  \nonumber \\	&=&  {5\over 4}q_{\rm s/v}^K(x_K) + r_{\rm s/v}^N(x) \left[  
  1 + {5+ \kappa\over 2}\,q_{\rm s/v}^K(x_K) + 
  {\kappa\over 4}\,R_{\rm s/u}^K(x_K) \right]
\label{eq:Rdef}
\end{eqnarray}
\normalsize
Measuring these cross sections requires a sufficiently intense 
and reasonably pure kaon beam.  
With the FNAL MI upgrade and a dedicated kaon beam line, one could 
envision substantial improvement in kaon flux, 
but this may still be below the rates
necessary to get precision DY measurements.

To estimate the magnitudes of expected DY ratios, we used kaon valence 
quark structure functions from Shigetani, Suzuki and 
Toki \cite {sst}.  They used a Nambu-Jona Lasinio [NJL] model to generate
the valence quark distributions.  These structure functions were able
to reproduce the CERN NA3 measurements of Badier et al.\ \cite{na3}, who 
compared kaon and pion-induced DY cross sections.  They showed that 
at large $x$ for the meson, the up quark structure function in the kaon
was substantially smaller than that for the pion.  This was expected since
the heavier $s$ quark of the kaon should carry more of the momentum
than the lighter nonstrange quark.   

In Fig.\ 1, we show the predicted valence quark distributions for the
kaon.  The dotted curve is the quark distribution at the hadronic
scale $Q_0^2 =$ (0.5 GeV)$^2$, while the solid curve is the
valence distribution evolved to $Q^2 =$ 20 GeV$^2$.  From Fig.\ 1b,
we see that the strange quark distribution is substantially bigger
than the nonstrange quark distribution at large $x_K$, in agreement
with the experimental measurements of Badier et al.\ \cite{na3}.  
Since we know of no sea quark distributions for the kaon, we used the
phenomenological sea quark distributions for the pion from Sutton et 
al.\ \cite{sutton,hmrs}.  The nucleon quark distributions were the CTEQ(3M)
fits from the CTEQ collaboration \cite{cteq}.  All of these were
calculated for $Q^2 = 20$ GeV$^2$.  

In Fig.\ 2 we show the quantity $\Sigma^{KD}_{\rm v}$ versus $x_K$ for 
various values of nucleon $x$.  As was discussed previously, 
this function separates in $x$ and 
$x_K$, so that if we integrate $\Sigma^{KD}_{\rm v}(x, x_K)$ over all
$x$ we can extract the up quark valence distribution in the kaon.  
The NJL distributions of 
Shigetani et al.\ fall off monotonically in $x_K$ for fixed $x$.  

In Figs.\ 3 we show predictions for $R_{\rm s/v}$ (the ratio of 
$\sigma^{K^+ D}_{DY}$ to $\Sigma^{KD}_{\rm v}$, defined in Eq.\ \ref{eq:Rdef}), 
versus \ $x_K$, for three values of nucleon momentum fraction 
($x= $ 0.2, 0.3 and 0.4). At each value of 
$x$, we show curves corresponding to four different kaon sea quark
distributions; these are taken from the pion sea quark  
distributions of Sutton et al. \cite{sutton}.  Those correspond to 
different fits to the NA10 Drell-Yan data \cite{naten,naelv}, where 
the meson sea carries from 5\% to 
20\% of the pion's momentum at $Q^2 = 20$ GeV$^2$ (i.e., these are fits
$2-5$ of Ref.\ \cite{sutton}).  The predicted ratio $R^K_{\rm s/v}$ is 
quite large: e.g., for $x= x_K = 0.2$, $R^K_{\rm s/v}$ varies from around 
0.2 to 0.5 depending on the momentum fraction carried by the sea. 
Furthermore,  
$R^K_{\rm s/v}$ is quite sensitive to the momentum fraction carried by the 
sea.  The quantity $R^K_{\rm s/v}$ is more or less linear in this momentum 
fraction; the difference between a parton distribution where 5\% of the 
kaon's momentum is carried by the sea, and one where 20\% of the momentum 
is carried by sea quarks, is roughly a factor of 3 in $R^K_{\rm s/v}$.  

Thus, assuming the availability of sufficiently intense separated beams
of kaons, even qualitative measurements of $R^K_{\rm s/v}$ would be able to
differentiate between kaon parton distributions where the sea carries
different fractions of the kaon's momentum. The process could be
measured at relatively small values of $x$ and $x_K$ provided 
that sufficiently
large count rates of muon pairs could be obtained. 

One remaining quantity is the kaon's strange quark valence distribution. 
In Eq.\ \ref{eq:Rdef} this arises from annihilation between a strange
valence quark in the kaon and the strange sea of the nucleon.  At
large $x_K$, the strange quark in the kaon should carry more of the
kaon's momentum than the nonstrange quark, due to the larger
mass of the strange quark.  This was confirmed in the NA3 measurements
of Badier et al.\ \cite{na3}, which showed that the $\bar{u}$ 
distribution in the $K^-$ was roughly half the $\bar{u}$ distribution
in the $\pi^-$ at large $x$, as measured in meson-nucleus Drell-Yan
processes.  This large-$x$ depletion of the nonstrange valence 
distribution in the kaon is expected to be offset by an increase 
in the strange quark valence distribution.  

In Figs.\ 4 we show the quantity $R^K_{\rm s/v}$ versus \ $x_K$, for 
two values
of the nucleon momentum fraction ($x = 0.2$ and 0.3).  The solid curves
include the kaon strange valence quark distribution; in the dashed
curves this distribution is set to zero (for illustrative purposes, 
to show the magnitude of the strange valence contribution).  
For large $x_K$, the
differences range from 40\% at small $x$ ($x \sim 0.2$), to about
15\% effects at larger $x \approx 0.4$.  However, the quantity 
$R^K_{\rm s/v}$ is small in this region, and it would be extremely
difficult to extract any information about the kaon's strange valence
quark distribution from such measurements.  This quantity is relatively
small due both to the charge (1/3) of the strange quark, and the 
strange/nonstrange ratio ($\kappa$) of the nucleon sea.  

We have assumed the validity of charge symmetry in this work.  From
previous investigations of charge symmetry violation [CSV] in such systems
(\cite{llrt,rtl,sather}) one would expect CSV effects to be at most
a few percent in the valence distributions.     
EMC-type effects (\cite{emc}) would also be expected to be small
for light isoscalar targets.  

In conclusion, with presently available kaon beams precision Drell-Yan
experiments are probably not feasible.  However, the CERN NA3 
group has already been able to obtain Drell-Yan cross sections 
integrated over $x$, and to extract qualitative results comparing 
pion and kaon-induced reactions.  It should also be possible for 
experiments to obtain interesting information on kaon structure 
at small $x$ and $x_K$.   


This work was supported by the Australian Research Council.
One of the authors [JTL] was supported in part by the US NSF under research 
contract NSF-PHY94-08843, and wishes to thank G.T. Garvey for useful
discussions.

\newpage

{\large {\bf Figure captions}}

\vspace*{0.5cm}

\normalsize

{\bf 1.} Theoretical kaon valence quark distributions, vs.\ Bjorken 
$x$, calculated using the NJL model of Shigetani et al., Ref.\ \cite{sst}.  
Dotted curve: quark distributions at the hadronic
scale, $Q_0^2 =$ (0.5 GeV)$^2$; solid curve: quark distributions
evolved to $Q^2 =$ 20 GeV$^2$.  (a) up quark valence distribution 
in the kaon, $x\,u^K_{\rm v}(x)$; (b) strange quark valence distribution 
in the kaon, $x\, s^K_{\rm v}(x)$.  

{\bf 2.} Predicted valence quark distributions for kaon-nucleus
reactions on isoscalar targets.  The quantity $\Sigma^{KD}_{\rm v}$ of 
Eq. (\ref{eq:sigmav}) is plotted vs.\ $x_K$ for four different values
of nucleon momentum fraction $x$.   
Solid curve: $x =0.6$; dashed curve: $x = 0.4$; long-dashed
curve: $x = 0.3$; dot-dashed curve: $x = 0.2$. Nucleon parton 
distributions are CTEQ(3M) parameterization of Ref.\ \cite{cteq}; kaon
valence distributions are those of Shigetani et al., Ref.\ \cite{sst}.   

{\bf 3.} Predicted sea/valence term $R^K_{\rm s/v}$ of Eq.\ \ref{eq:Rdef}, 
vs.\ the kaon momentum fraction $x_K$, for various kaon sea quark
distributions, which vary according to the fraction of the kaon's 
momentum carried by the kaon sea.  Solid curve: kaon sea carries 20\% of 
the kaon's momentum; dashed curve: kaon sea of 15\%; 
long-dashed curve: kaon sea of 10\%; dot-dashed curve: kaon sea of 
5\%.  (a) Nucleon momentum fraction $x = 0.2$; (b) $x = 0.3$; (c)
$x = 0.4$.  Valence parton distributions are those of Fig.\ 1.  Meson
sea quark distributions are pion sea quark distributions of Sutton et al., 
Ref.\ \cite{sutton}. 

{\bf 4.} Dependence of sea/valence ratio $R^K_{\rm s/v}$ on the 
strange valence distribution of the kaon, plotted v.\ kaon momentum fraction
$x_K$.  Solid curve: the quantity 
$R^K_{\rm s/v}$ of Eq.\ \ref{eq:Rdef} including the kaon strange
valence distribution; dashed curve: the same quantity where the kaon
strange valence distribution is set to zero. (a) Nucleon momentum fraction
$x =0.2$; (b) nucleon momentum fraction $x =0.3$.   
Nucleon and kaon parton distributions are 
those of Fig.\ 2.    
\\


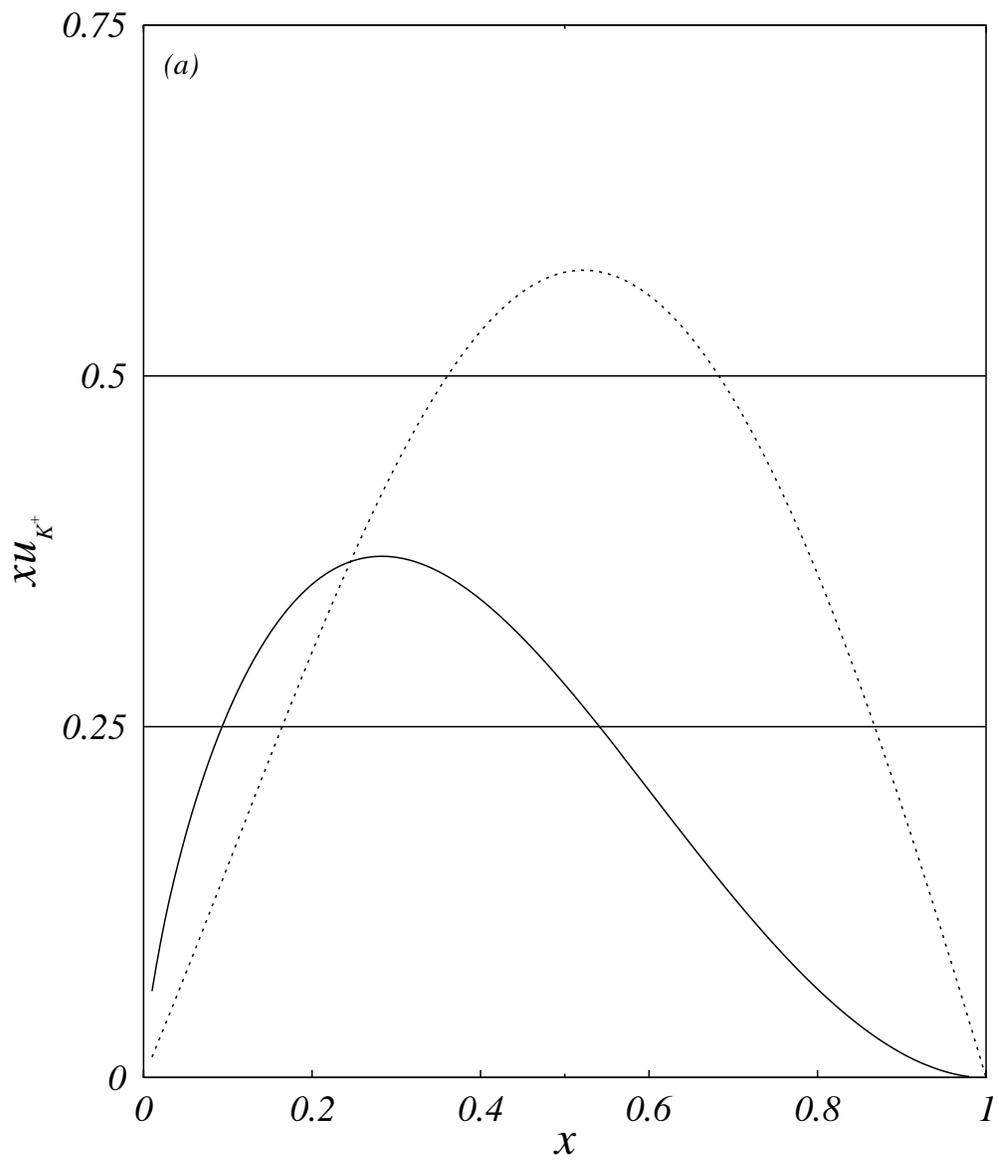
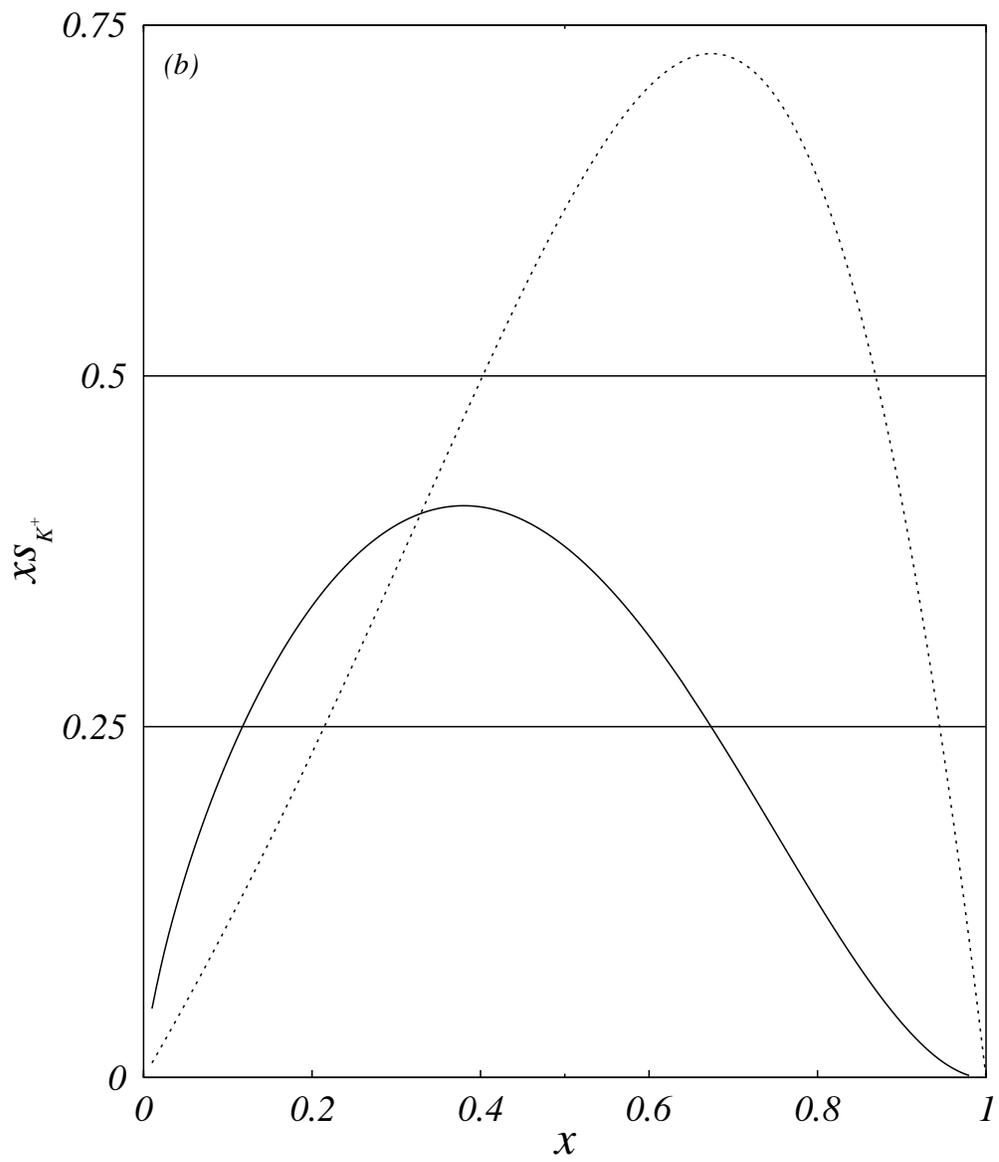
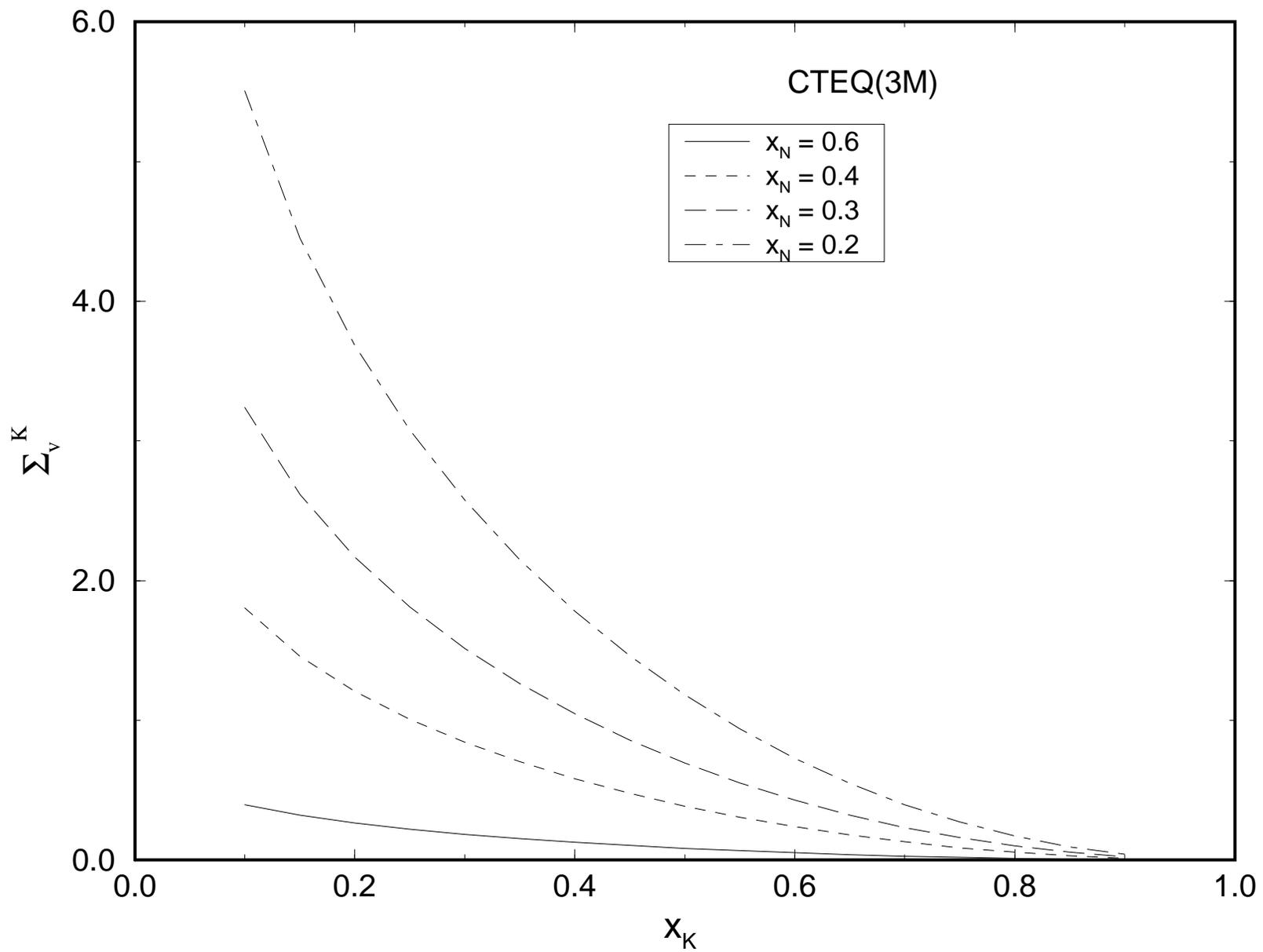
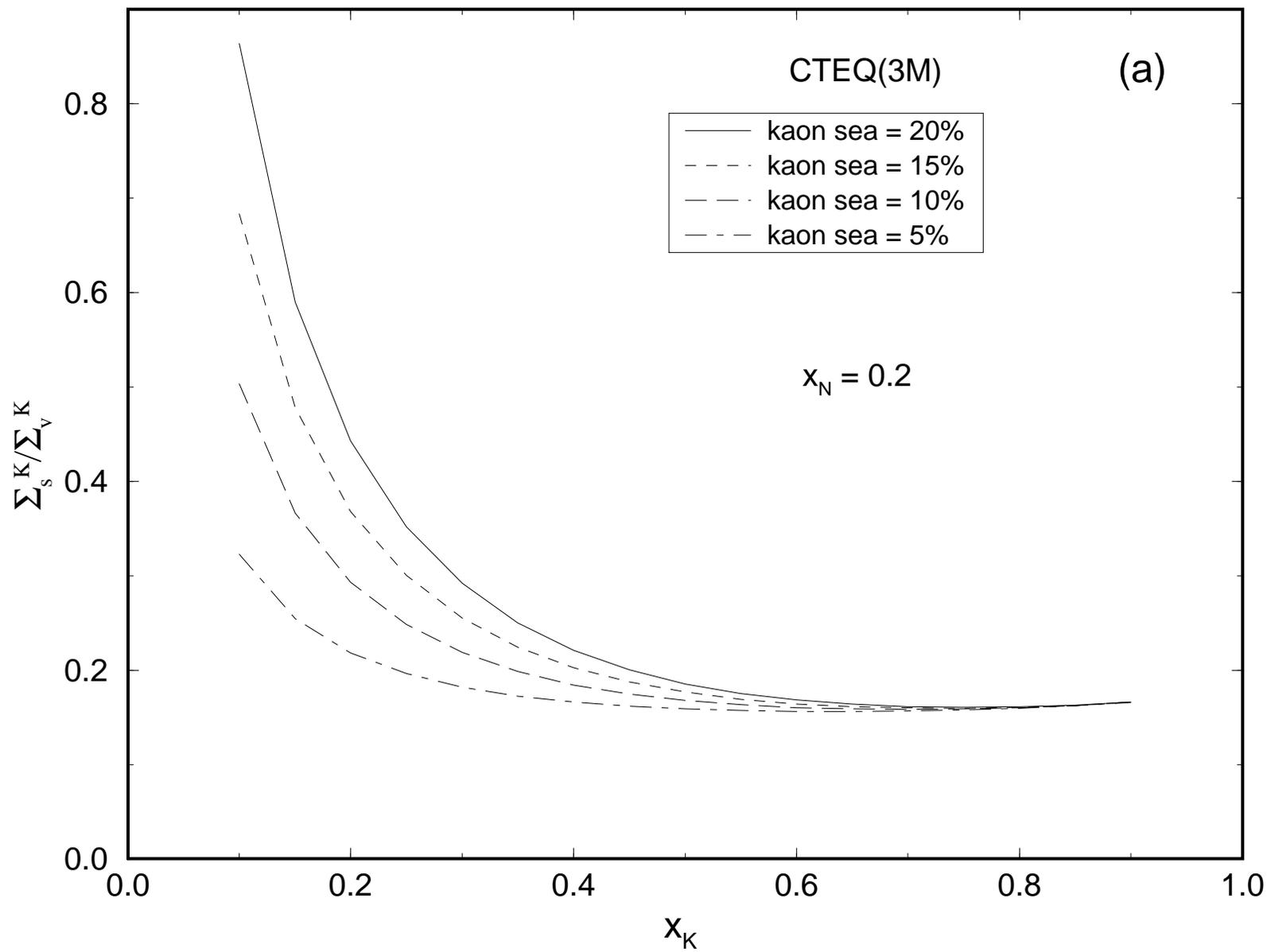
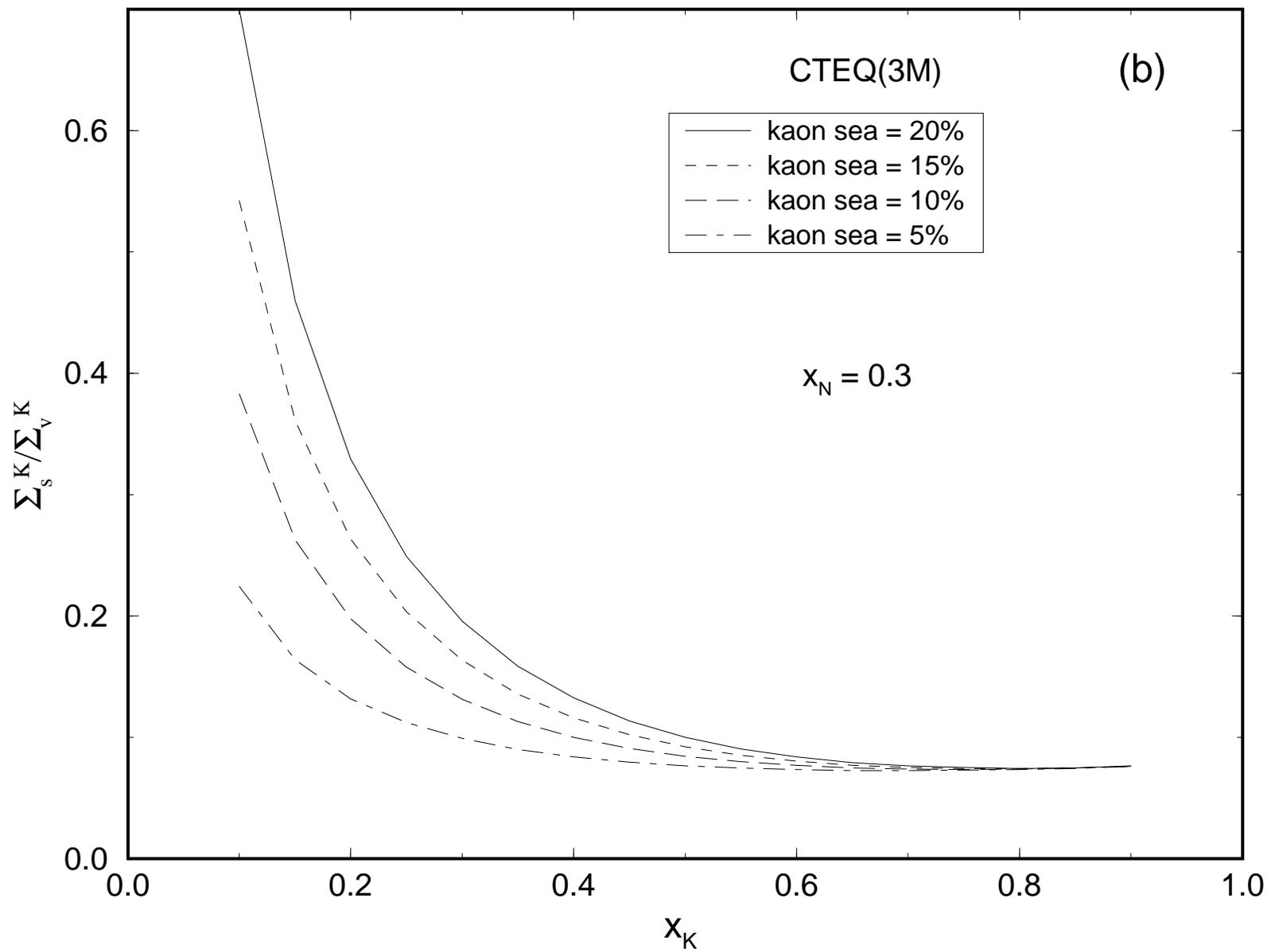
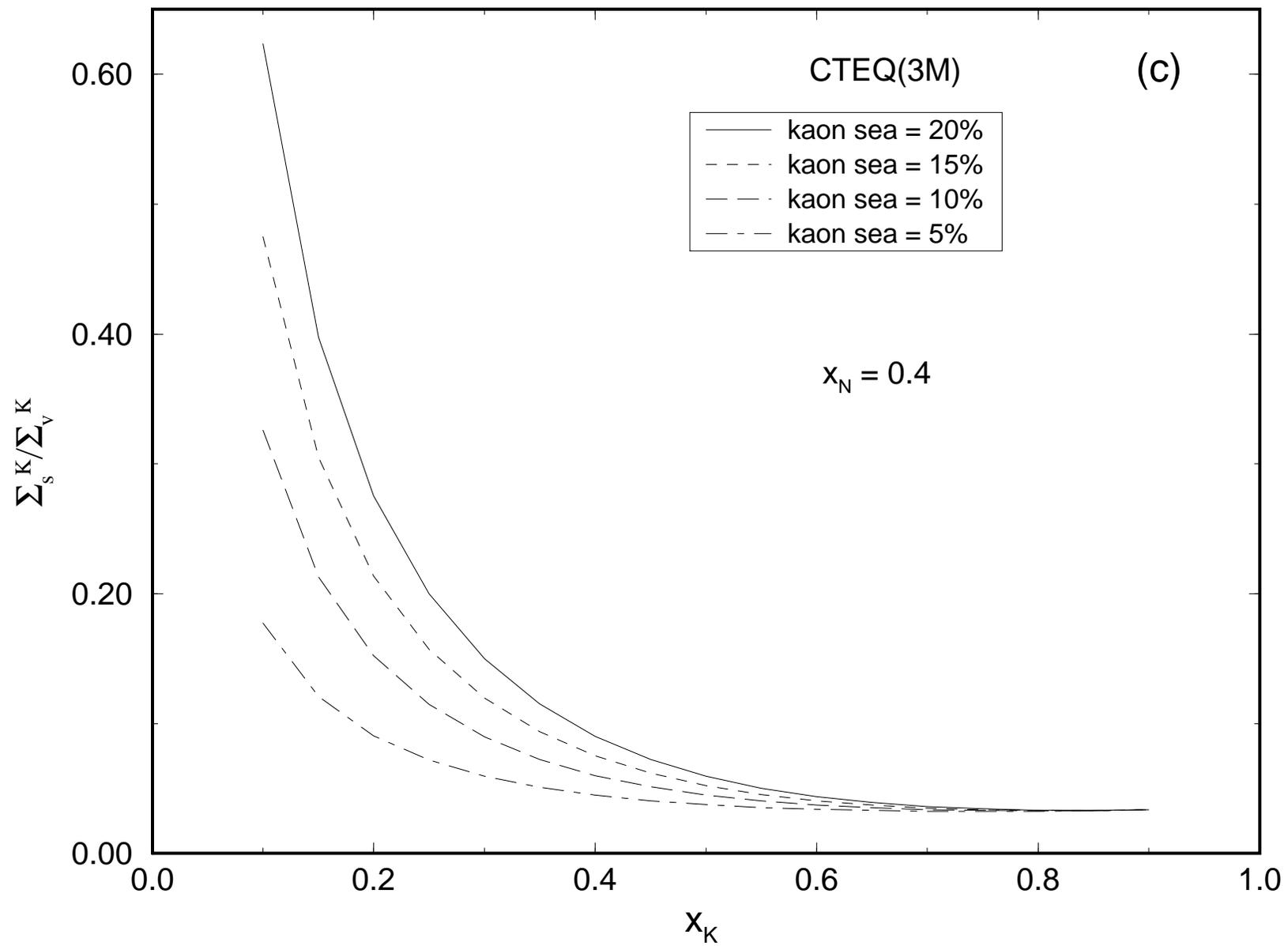
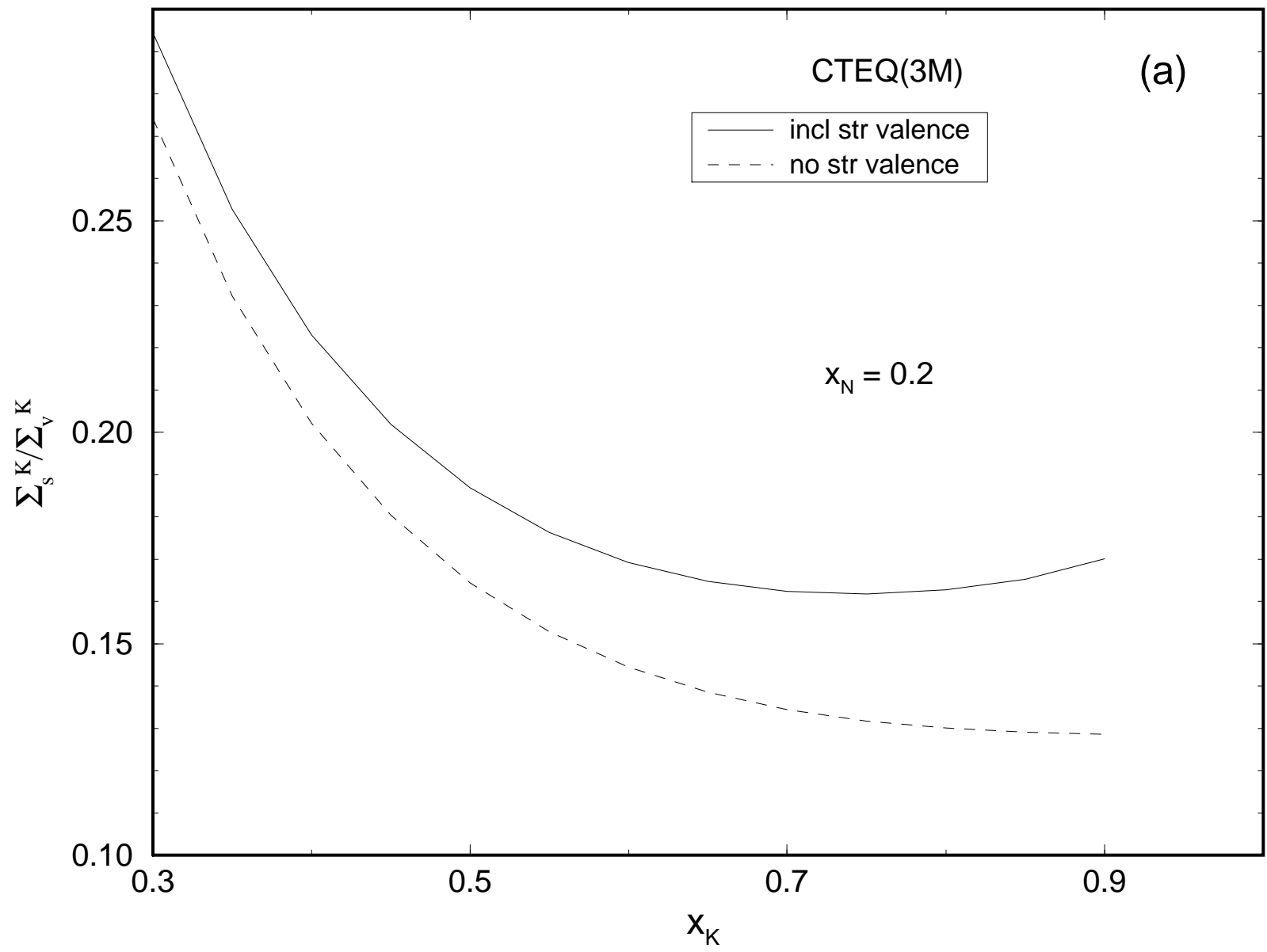
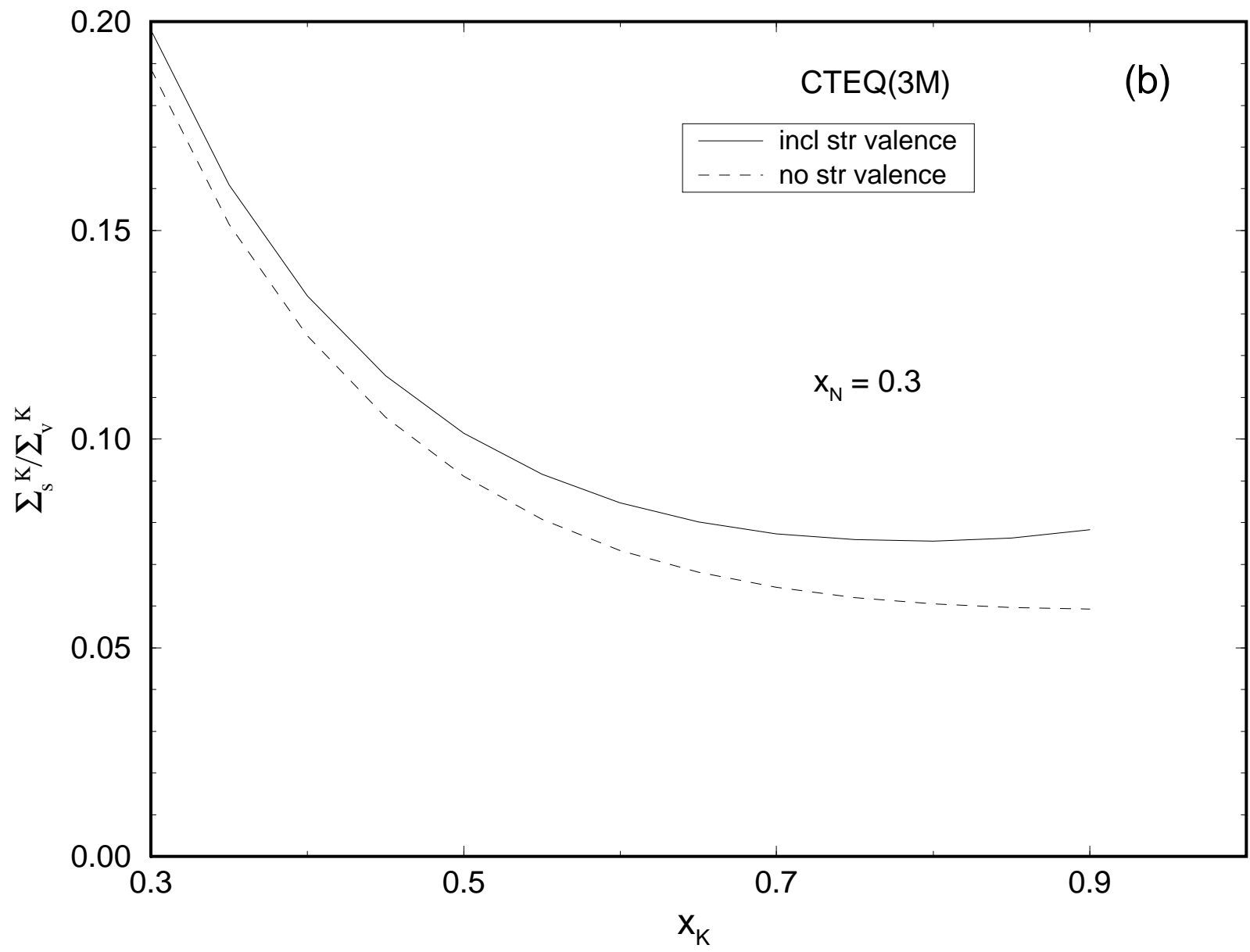
\end{document}